\documentclass[aps,floats,prd,nofootinbib,twocolumn]{revtex4-1}

\usepackage{amssymb}
\usepackage{amsmath}

\usepackage{graphicx}
\usepackage{graphics}
\usepackage{dcolumn}
\usepackage{color}
\usepackage{fancyhdr} 
\usepackage{graphicx}

\usepackage{bibshortcuts}

\usepackage[utf8]{inputenc}

\usepackage{subcaption}
\usepackage{ragged2e}
\DeclareCaptionJustification{justified}{\justifying}
\def\VEV#1{\left\langle #1 \right\rangle}

    \newcommand{\be}{\begin{equation}}
  \newcommand{\ee}{\end{equation}}
    \newcommand{\ba}{\begin{align}}
  \newcommand{\ea}{\end{align}}

\begin{document}

\title{Finding the Missing Baryons with Fast Radio Bursts \\ and Sunyaev-Zeldovich Maps}

\author{Julian B.~Mu\~noz\footnote{Electronic address: \tt julianmunoz@fas.harvard.edu}
} 
\affiliation{Department of Physics, Harvard University, 17 Oxford St., Cambridge, MA 02138}
\author{Abraham Loeb}
\affiliation{Astronomy Department, Harvard University, 60 Garden St., Cambridge, MA 02138}

\date{\today}

\begin{abstract}
Almost a third of the cosmic baryons are ``missing" at low redshifts, as they reside in the invisible warm-hot intergalactic medium (WHIM).
The thermal Sunyaev-Zeldovich (tSZ) effect, which measures the line-of-sight integral of the plasma pressure, can potentially detect this WHIM, although its expected signal is hidden below the noise.
Extragalactic dispersion measures (DMs)---obtained through observations of fast radio bursts (FRBs)---are excellent tracers of the WHIM, as they measure the column density of plasma, regardless of its temperature.
Here we propose cross correlating DMs and tSZ maps as a new way to find and characterize the missing baryons in the WHIM.
Our method relies on the precise ($\sim$ arcminute) angular localization of FRBs to assign each burst a DM and a $y$ parameter.
We forecast that the signal from the WHIM should be confidently detected in a cross-correlation analysis of $\sim10^4$ FRBs, expected to be gathered in a year of operation of the upcoming CHIME and HIRAX radio arrays, confirming the recent tentative detections of filamentary WHIM.
Using this technique, future CMB probes (which might lower the tSZ noise) could determine both the temperature of the WHIM and its evolution to within tens of percent.
Altogether, DM-tSZ cross correlations hold great promise for studying the baryons in the local Universe.
\end{abstract}

\maketitle

\section{Introduction}

Data from both the epoch of recombination~\cite{Ade:2015xua}, and Big Bang nucleosynthesis~\cite{Cyburt:2015mya}, predict a larger baryonic energy density than observed so far in the local Universe.
This tension in the baryonic census has been dubbed the``missing-baryon problem"~\cite{Fukugita:1997bi,Shull:2011aa}, and has loomed as one of the longest unsolved puzzles in astrophysics.
A target for the location of these  baryons was identified in Ref.~\cite{Cen:1998hc}, where it was argued that a large fraction ($30-50\%$) of baryons can reside in a warm-hot intergalactic medium (WHIM) phase, with temperatures in the range $T_{\rm whim}=10^{5-7}$ K, too warm to absorb efficiently~\cite{Nicastro:2007gv,TepperGarcia:2010sf} and too cold to emit X-rays~\cite{Bregman:2009rz}.

Nonetheless, these baryons could be observed through their Sunyaev-Zeldovich (SZ) signature in the cosmic microwave background (CMB)~\cite{Sunyaev:1970eu,Ho:2009iw,HernandezMonteagudo:2006kv,Genova-Santos:2013qfa}, although detection is hindered by the much-larger signal from collapsed regions, such as galaxy clusters~\cite{Seljak:2001rc,Battaglia:2010tm}.
One way to overcome this difficulty is by cross correlating thermal SZ (tSZ) maps with tracers of the WHIM~\cite{Atrio-Barandela:2017atd,VanWaerbeke:2013cfa,Hernandez-Monteagudo:2015cfa}, although care must be exercised to avoid contamination from clusters~\cite{Battaglia:2014era}.
In this spirit, two groups have recently reported a positive correlation between galaxy pairs (acting as tracers of intergalactic gas filaments) and tSZ maps, consistent with the missing baryons in the WHIM~\cite{Tanimura:2017ixt,deGraaff:2017byg}.
This is the first claimed detection of its kind which, if confirmed, can account for the entirety of the missing baryons in the local Universe.
It is, therefore, essential to reproduce this measurement with a different tracer of the WHIM.
Here, we propose cross correlating tSZ maps with extragalactic dispersion measures (DMs), obtained from observations of fast radio bursts (FRBs), as a way to probe the WHIM and find the missing baryons.

FRBs are short (ms) and bright (Jy) radio (GHz) transients~\cite{Lorimer:2007qn}.
Their DM directly measures the integrated column density of free electrons along the line of sight.
A few dozen FRBs have been detected to date~\cite{Petroff:2016tcr}, all with DMs in excess of Galactic expectations~\cite{Cordes:2002wz,YMW16}, signaling an extragalactic origin (with typical inferred redshift $z\sim0.5$).
While their sources remain elusive, it is clear that FRBs can be powerful beacons for cosmology. 
They have been proposed as a dark-matter detector through gravitational lensing~\cite{Munoz:2016tmg}, for improving measurements of cosmological parameters~\cite{Walters:2017afr,Yang:2016zbm}, finding circumgalactic baryons~\cite{McQuinn:2013tmc,Fujita:2016yve,Ravi:2018ose}, or detecting Helium reionization~\cite{Zheng:2014rpa}.

Here we develop a cross-correlation technique for tSZ maps and DMs, which can determine if the WHIM hosts the missing baryons, as well as constrain its thermal state.
Measurements of the tSZ effect~\cite{Sunyaev:1970eu}, expressed through the Comptonization parameter $y$, are most sensitive to the densest and warmest regions of the Universe.
Extragalactic DMs, on the other hand, simply trace gas along the line of sight.
Correlating these two probes can, thus, unearth the WHIM signature by underweighting the hottest regions (such as galaxy clusters), while overweighting the average-density intergalactic medium (IGM).
Operationally, we propose performing the $y$-DM cross correlation on an event-by-event basis.
Each detected FRB will be localized within arcminutes in the sky~\cite{Masui2017}, a scale comparable to the width of CMB $y$ pixels~\cite{Aghanim:2015eva,Abazajian:2016yjj}.
We can, then, assign each FRB an observed DM, and a $y$ parameter, and find if these two quantities are correlated.
A detection of DM-$y$ correlation, or a lack thereof, can confirm whether the WHIM detected in Refs.~\cite{Tanimura:2017ixt,deGraaff:2017byg} accounts for all the missing baryons.
Additionally, we will argue that the time evolution of this correlation can map out the thermal history of baryons in the low-redshift Universe.

Our paper is structured as follows. In Sections~\ref{sec:DMs} and~\ref{sec:tSZ} we describe our two observables: extragalactic DMs and  tSZ maps. We outline the cross-correlation method in Section~\ref{sec:Method}, and study a few applications in Section~\ref{sec:App}. We conclude in Section~\ref{sec:Conclusions}.
Throughout this discussion, we will assume standard $\Lambda$CDM cosmology, with a Hubble constant of $H_0=70$ km s$^{-1}$ Mpc$^{-1}$, and baryon and CDM densities of $\Omega_b=0.045$ and $\Omega_c=0.24$, consistent with the latest Planck data~\cite{Aghanim:2018eyx}.

\section{Dispersion Measures from \newline Fast Radio Bursts}
\label{sec:DMs}

We begin by reviewing extragalactic DMs from fast radio bursts.
For a recent review of FRBs see Ref.~\cite{Katz:2018xiu}, and Ref.~\cite{Petroff:2016tcr} for a catalog of all observed FRBs to date.
The origin of FRBs remains uncertain, with many potential candidate sources, ranging from merging white dwarfs~\cite{Kashiyama:2013gza}, to neutron stars~\cite{Cordes:2015fua}, including young magnetars~\cite{Metzger:2017wdz}, and light sails~\cite{Lingam:2017fwa}.
Regardless of their origin, these short bursts hold great promise for cosmological studies, as they are expected to occur often in our Universe~\cite{Fialkov:2017qoz}, and to be detected in the tens of thousands by the upcoming CHIME~\cite{ChimeFRB} and HIRAX~\cite{Newburgh:2016mwi} radio arrays, as well as potentially by HERA~\cite{DeBoer:2016tnn} at lower frequencies.

Similarly to radio emission from pulsars, FRBs are dispersed by intervening gas.
A plasma of free electrons causes a delay of low-frequency electromagnetic waves traveling through it, as the dispersion relation of these waves is modified by the presence of the plasma frequency, 
\be
\omega_{\rm pl} = \sqrt{\dfrac{2 \alpha\,  h \, c \, n_e}{m_e}},
\ee
with a value of $\omega_{\rm pl} = 27$ s$^{-1}\times(1+z)^{3/2}$ for the average-density IGM,
where $m_e$ is the electron mass, $n_e$ is the number density of free electrons, $c$ is the speed of light, and $\alpha$ and $h$ are the fine-structure and Planck constants, respectively.
This gives rise to an effective frequency-dependent group velocity
\be
\dfrac{v_g(\omega)}{c} \approx 1 -  \dfrac{\omega_{\rm pl}^2}{2  \omega^2},
\ee
for $\omega \gg \omega_{\rm pl}$,
which causes lower-frequency signals to arrive later.
This effect is commonly  characterized through a dispersion measure, defined for Galactic sources as
\be
{\rm DM} = \int ds\, n_e(s),
\label{eq:DMdef}
\ee
where $s$ is a physical distance.
We will divide the DM of any FRB into two components, one due to electrons in the Milky Way (disk and halo) plus the host galaxy (including any contribution from the source itself), and the other from the intervening IGM. 

\subsection{Galactic and Host Contributions}

The DM defined in Eq.~\eqref{eq:DMdef} is routinely measured in pulsars, for which interstellar gas in the Milky Way disk, and halo, generates a dispersion measure dependent on the pulsar Galactic coordinates~\cite{Cordes:2002wz}.
High-latitude pulsars show dispersion measures of order $\sim 50$ pc cm$^{-3}$, whereas the ones towards the galactic plane can show DMs in excess of $ 10^3$ pc cm$^{-3}$, owing to the larger electron density in the disk.
This contribution can, however, be subtracted with relatively small uncertainties, given our knowledge of the electron distribution in the Milky Way~\cite{Cordes:2002wz,YMW16}, so we will ignore it for the rest of this work.

A larger uncertainty comes from the unknown contribution to the DM from to the FRB source and its host galaxy.
If the host galaxy is anything like our own Milky Way, its shape and orientation can have a large impact on the observed DM~\cite{Yang:2017bls}.
Recently, one of the detected bursts, FRB 121102, has been located within a
dwarf galaxy at $z=0.19$~\cite{Chatterjee:2017dqg}.
Given the (cosmological) distance to this galaxy, the host+source contribution to the DM has been estimated to be in the range
\be
 55 \leq  \dfrac{ \left .{\rm DM}\right._{\rm host+source}}{\rm pc\, cm^{-3}} \leq 225
\ee
for this FRB~\cite{Tendulkar:2017vuq}, accounting for at most a third of the observed DM. 
Thus, the host contribution is not expected to compose the majority of the observed DM for $z\gg0.1$.

We will not attempt to model the contribution from each host in detail, and instead assume that---once angular-averaging over all host configurations, and sizes, is taken into account---the host + source contribution to the DM for an FRB from redshift $z$ is simply given by
\be
\overline{\rm DM}_{\rm host} (z)= \dfrac{ \overline{\rm DM}_h}{1+z},
\label{eq:DMhost}
\ee
where the factor of $(1+z)^{-1}$ arises because of time dilation, and we adopt a constant normalization factor $\overline{\rm DM}_h  = 100 $ pc cm$^{-3}$ as an approximation (ignoring host evolution in redshift). 
Moreover, we will assume that the possible values of ${\rm DM}_{\rm host}$ for every FRB are distributed uniformly between 0 and $2\,\overline{\rm DM}_{\rm host} (z)$ for each FRB, to account for the relative orientations of the FRB host and source.
While this is an overly simplistic model, we will see that it does not affect our results considerably.

\subsection{IGM Contribution}

The DM component that we are interested in is that of the IGM.
To compute it, we need to extend Eq.~\eqref{eq:DMdef} for extragalactic sources, accounting for the effects of cosmological redshift, as in Ref.~\cite{Zheng:2014rpa}.
We start by calculating the delay of a signal with observed frequency $\omega_{\rm obs}$ from redshift $z$,
\be
\Delta t (z, \omega_{\rm obs}) = \dfrac 1 {2c} \int_0^z dz' \dfrac{ds}{dz'}  \dfrac{\omega_{\rm pl}^2(z')}{\omega^2(z')} (1+z'),
\label{eq:Deltat}
\ee
where $ds = c\, dt$, the last factor of $(1+z')$ is due to time dilation between the origin of the delay and us, and $\omega(z) = \omega_{\rm obs} (1+z)$, due to the redshifting of photons.
The conversion factor from physical distance to redshift is simply $ds/dz = c \, (1+z)^{-1} H^{-1}(z)$, so substituting in Eq.~\eqref{eq:Deltat}, and remembering that DM $\propto \Delta t(\omega_{\rm obs})\, \omega_{\rm obs}^2$, we obtain
\be
\left.{\rm DM}\right. _{\rm IGM}(z) = n_e^{(0)} c \int_0^z \dfrac{dz' (1+z')}{H(z')},
\label{eq:DMz}
\ee
where we assumed that both Hydrogen and Helium are fully ionized, as expected at $z\lesssim3$, so the number density of electrons behaves as $n_e(z) = n_e^{(0)} (1+z)^3$, where $n_e^{(0)}$ is the number density of electrons today, given by
\be
n_e^{(0)} = \dfrac{\Omega_b \rho_{\rm crit}}{m_H} (1-Y_{\rm He}) (1+2 f_{\rm He}) \approx 2.2\times 10^{-7}\,{\rm cm}^{-3},
\label{eq:ne0}
\ee
for our fiducial parameters, where $m_H$ is the hydrogen atom mass, and 
where we set $Y_{\rm He}=0.24$, yielding $f_{\rm He} \equiv n_{\rm He}/n_H = 0.08$.
For reference, we have found that 
we can approximate
\be
\left.{\rm DM}\right._{\rm IGM}(z)  \approx 1.1\, \dfrac{c\, n_e^{(0)}}{H_0} z \equiv \overline{\rm DM} \times z,
\label{eq:DMzlinear}
\ee 
for $z\lesssim2$,
with $\overline{\rm DM} = 1025$ pc cm$^{-3}$.
This value should be lowered by a factor of $(1-f_{\rm coll})$ for a fraction $f_{\rm coll}$ of  electrons residing in collapsed objects, which should be tracked separately.
For simplicity, we set $f_{\rm coll}=0$ in what follows although its value today is estimated at $\sim 5-10\%$~\cite{Deng:2013aga}.
The warm-hot intergalactic medium (WHIM), composing a fraction $f_{\rm whim}$ of the baryons, would account for part of the DM in Eq.~\eqref{eq:DMz}, while the rest of it is sourced by the cold IGM. 
These two components are, however, highly correlated, so we will use DM$_{\rm IGM}$ as a measure of the amount of gas along the line of sight towards each FRB source.

Fluctuations in the IGM density produce variations in DM$_{\rm IGM}$ between different lines of sight at the ten-percent level~\cite{McQuinn:2013tmc}.
We will, however, not model this effect, since any gas fluctuation---sourcing both a change to DM and $y_{\rm whim}$---can be reabsorbed into a different FRB redshift, effectively included in the results that we calculate.
Additionally, while from Eq.~\eqref{eq:ne0} it might appear that one can measure the baryon abundance ($\Omega_b$) to great precision simply with DMs (without tSZ information), we note that this requires knowledge of the redshift of each FRB, and is degenerate with the largely unknown host+source contribution to the DM~\cite{Walker:2018qmw}.
The method we propose is immune to these uncertainties.

\section{Thermal Sunyaev-Zeldovich effect}
\label{sec:tSZ}

Along their path to Earth, cosmic-microwave-background (CMB) photons may change their energies through inverse Compton scattering on hot electrons, giving rise to the thermal Sunyaev-Zeldovich (tSZ) effect~\cite{Sunyaev:1970eu}. 
The resulting change in the CMB temperature is given by
\be
\dfrac{\Delta T(\hat n, \nu)}{T_{\rm CMB}} = y(\hat n) g(\nu),
\label{eq:ydef}
\ee
where $T_{\rm CMB}$ is the average CMB temperature, $g(\nu) = x \coth(x/2) -4$, with $x \equiv h\nu/(k_B T_{\rm CMB})$, where $k_B$ is the Boltzmann constant, and we use the standard definition of the Comptonization parameter,
\be
y (\hat n) = \dfrac{ \sigma_T k_B}{m_e c^2} \int d s\,  n_e T_e,
\ee
where $\sigma_T$ is the Thomson cross section, $T_e$ is the electron temperature, and $ds$ is the proper line-of-sight element.

In addition to the tSZ effect that we will focus on, the bulk velocity of electrons produces a kinematic SZ (kSZ) effect, which can also be used to search for the missing baryons~\cite{Ho:2009iw}. 
This effect was first detected in Ref.~\cite{Hand:2012ui}, and subsequent observations have utilized it to confirmed that the amount of baryons in the local Universe is in agreement with BBN and CMB expectations~\cite{Ade:2015lza,Schaan:2015uaa,Hill:2016dta,Ferraro:2016ymw}. 
Nonetheless, kSZ measurements do not provide insights on the thermal state of the missing baryons, whereas tSZ maps are imprinted with that information.

Most of the cosmic tSZ signal is sourced by galaxy clusters, where electrons are virially heated to temperatures of $\sim 1-10$ keV~\cite{Aghanim:2015eva,Dunkley:2010ge,Komatsu:2010fb}.
Nonetheless, the WHIM is expected to compose a fraction ($\sim 15\%$) of the total tSZ signal in the local Universe~\cite{HernandezMonteagudo:2006kv}, albeit being fairly diffuse across the sky (as opposed to the concentrated signal from clusters).
We will use DMs as tracers of intergalactic gas to unearth the WHIM signal in tSZ maps.

Given the distinct spectral dependence of the tSZ effect in Eq.~\eqref{eq:ydef}, the best way to extract the $y$ parameter from CMB observations is to add  the information implicit in different frequency maps~\cite{Dunkley:2010ge}. 
In particular, the Planck collaboration performed an internal linear combination (ILC) algorithm of six of their channels to obtain a $y$-map of the sky~\cite{Aghanim:2015eva}.
We will assume a foreground-cleaned $y$ map, from where one can decompose the observed $y_{\rm obs}$ parameter in every pixel as
\be
y_{\rm obs} (\hat n) =y_{\rm IGM} (\hat n) + y_{\rm cl} (\hat n)  + y_{\rm noise} (\hat n),
\ee
where the first two components correspond to the IGM and to clusters, respectively, and the last component is noise. This decomposition will, however, not be valid if the $y$ maps are contaminated with foregrounds, which can induce correlations between all these components.
Unless otherwise stated, we will assume that $y_{\rm noise}$ is given by a Gaussian distribution, which includes both instrumental and confusion noise, and we will reevaluate this assumption in Section~\ref{sec:Method}.
For now, we focus on the first two terms.

\subsection{Warm-Hot Intergalactic Medium}

The average temperature of the IGM prior to shock heating via structure formation is around 1 eV ($\sim 10^4$ K), as set by the competition between photoheating and adiabatic expansion~\cite{Sanderbeck:2015bba}, far too cold to be observable in tSZ maps.
However, shocks during structure formation at low redshifts heat up a large fraction of the gas (the WHIM) to temperatures around $\sim 10^6$ K~\cite{Cen:1998hc,Dave:2000vh}. 
Different methods have been proposed to separate this WHIM in tSZ maps from the much-larger signal originating in clusters~\cite{Goldberg:1999xm,Atrio-Barandela:2017atd,Genova-Santos:2013qfa,Ma:2014dea}.
By cross correlating tSZ and lensing maps, a detection of the WHIM was reported in Ref.~\cite{VanWaerbeke:2013cfa}.
Later studies, however, found that the observed tSZ-lensing correlation arises naturally through correlations in the intracluster medium~\cite{Battaglia:2014era}, and does not require the presence of a WHIM.

Interestingly, two groups have recently reported a 5-$\sigma$ detection of tSZ emission from filamentary gas between galaxies~\cite{Tanimura:2017ixt,deGraaff:2017byg}. 
To obtain their measurements, these groups cross correlated tSZ maps with positions of known galaxies, from the Sloan Digital Sky Survey, and added the $y$ signal between all nearby galaxy pairs (using the LRG~\cite{Tanimura:2017ixt} and  CMASS~\cite{deGraaff:2017byg} galaxy catalogs).
This detection, however, can only account for gas in filaments between observed galaxies, and could be easily contaminated by hot gas in the vicinity of galaxies.
We will show how DMs, as tracers of gas, provide us with direct way to probe the WHIM, which will be able to confirm these tentative detections.

The WHIM component of the tSZ effect, denoted by the $y_{\rm whim}$ parameter, can be written as~\cite{Seljak:2001rc}
\be
y_{\rm whim} (z) = f_{\rm whim}  \dfrac{k_B \sigma_T}{m_e \, c \, } n_e^{(0)} \int_0^{z}\!\! dz'\dfrac{ T_{e}(z') (1+z')^2}{H(z')},
\label{eq:ywhim}
\ee
where  $f_{\rm whim}$ is the fraction of baryons that are in the WHIM, and we ignore a possible $y$ component coming from non shock-heated regions (with $T<10^5$ K), as this has been shown to compose a negligible part of the total tSZ luminosity~\cite{HernandezMonteagudo:2006kv}.
Comparing this equation to Eq.~\eqref{eq:DMdef} we see how both DM$_{\rm IGM}$ and $y_{\rm whim}$ trace the same underlying gas distribution, albeit $y_{\rm whim}$ also depends on its temperature and fraction $f_{\rm whim}$ of all baryons.

The thermal state of the WHIM, modeled through $f_{\rm whim}$ and $T_{e}(z)$, determines the size of the DM-$y$ cross correlation.  
To exemplify this, we will consider two models of the WHIM:  

$\bullet$ Model I:  With $f_{\rm whim}=1$ and $T_{e}(z) = 100\, (1+z)^{-1}$ eV, as indicated by the simulations of Ref.~\cite{Goldberg:1999xm}.

$\bullet$ Model II:  With $f_{\rm whim}=0.3$ and electron temperature $T_{e}^{(0)} = 10^6$ K today, redshifting as $e^{-3\,z/2}$, where the exponent is fitted to the simulations of Ref.~\cite{Cen:1998hc}.

We will use Model I, due to its simplicity, as a benchmark to determine the detectability of the signal with current experiments and we will use the realistic Model II to forecast how well future experiments can constrain the thermal state of the WHIM.

\subsection{Clusters and Galaxies}

In addition to the sought-after WHIM-induced DM-$y$ cross correlation, intervening galaxies and clusters can generate a different DM-$y$ cross correlation. 
We will now estimate this latter contribution and discuss how to mitigate it.

We begin by estimating the probability that an FRB intersects a galaxy cluster.
Given a cluster comoving number density of $n_{\rm cl}\approx 10^{-5}\,h^3$ Mpc$^{-3}$, and a virial radius of $R\sim 1\,h^{-1}$Mpc, the probability that an FRB from a redshift $\overline z_s\approx 0.5$ (at the median of the FRB distribution) crosses a cluster on its way to us is $\sim 4\%$.
Considering that every cluster contributes with an average $y_{\rm cl}\sim10^{-6}$~\cite{Komatsu:2002wc,Horowitz:2016dwk} which we also take to be redshift independent,
we find that every FRB sees a cluster contribution to the $y_{\rm obs}$ of
\be
\left. y_{\rm cl}\right|_{\rm FRB} \sim 4 \times 10^{-8},
\label{eq:ycluster}
\ee
which is comparable to the WHIM signal we are after, which has an integrated strength out to $z=0.5$ of $y_{\rm whim} \sim \{20,5\} \times 10^{-8}$, for thermal models \{I,II\}.
Note, however, that the cluster number density quickly decreases for $z\gtrsim 0.5$~\cite{Carlstrom:2002na}, so higher-$z$ FRBs will have a comparatively smaller cluster imprint in the DM-$y$ correlation. For instance, we find, using the Sheth-Tormen mass function~\cite{Sheth:1999su}, that a source (FRB) from $z=0.5$ would cross within the virial radius of a cluster an average of one in a hundred times, somewhat lower than our simple estimate above.

A similar DM-tSZ correlation can appear due to FRBs that originate within clusters, which would therefore be co-located with them.
The likelihood of a typical FRB host galaxy to reside in a cluster depends on its color, magnitude, and other properties~\cite{Fialkov:2017vsq}.
For bright galaxies as FRB hosts, we expect 5\% of those to reside in rich clusters~\cite{Bahcall:1995tf}.
This would be comparable to the probability of crossing a cluster that we estimated above, so the same caveats would apply.

We now discuss three ways to mitigate the effect of clusters in the DM-$y$ cross correlation.
First, FRBs are expected to be scattered when crossing a cluster, due to turbulence in the intracluster medium (ICM). This could identify which FRBs have crossed a cluster~\cite{Macquart:2013nba,McQuinn:2013tmc}.
Second, given the high resolution of radio interferometers to locate FRBs~\cite{ChimeFRB}, and the sensitivity of current and upcoming CMB and spectroscopic surveys to detect and locate clusters~\cite{Abazajian:2016yjj,Wen:2012tm}, most FRBs coincident with a cluster can be separated off from the general population, which would significantly lower the $\left. y_{\rm cl}\right|_{\rm FRB}$ term.
These separated FRBs (either by localization or by scattering) could, however, be used to learn about the gas content of clusters and galaxies.
For instance, in Ref.~\cite{Fujita:2016yve} it was shown that with Square-Kilometer Array (SKA) observations of FRBs, one could detect the $y$-DM correlation of circumgalactic baryons in clusters beyond the virial radius.
Finally, we will show in Sec.~\ref{sec:App} how the expected DM-$y$ correlation from clusters scales differently from that of the WHIM, acting as a diagnostic of the origin of the signal.

Additionally, very massive galaxies and groups can also produce a tSZ signal. 
The typical contribution of a luminous red galaxy (LRG), with stellar mass $M_*\geq10^{11.3}M_\odot$ (corresponding to $M_{\rm halo}\gtrsim 10^{14} M_\odot$~\cite{Mitchell2016}), is estimated to be $y_{\rm gal}\sim10^{-7}$~\cite{Tanimura:2017ixt}.
The probability of an FRB from $z=0.5$ intersecting a halo with $M_{\rm halo}=10^{14}\,M_\odot$ is below 10\%~\cite{McQuinn:2013tmc}, yielding a contribution to the DM-$y$ correlation from these galaxies of $y_{\rm gal} \lesssim 10^{-8}$, safely below that of clusters.
Even though the majority of clusters and galaxies are unresolved in  contemporary tSZ maps, their signal increases $y_{\rm noise}$ through confusion noise and produces non-Gaussianities in $y$. 
We emphasize that this is unrelated to whether an FRB crosses these sources, and is included in our calculations.

Finally, we are ignoring any tSZ contribution from the host of the FRB, which could correlate with the DM$_{\rm host}$ component. Adding it would only increase the signal, as it would correlate positively with DM, although from the one FRB host known, a dwarf galaxy at $z=0.19$~\cite{Tendulkar:2017vuq}, we do not expect any measurable tSZ signal.
Given the typical Compton parameter $y_{\rm gal} \lesssim 10^{-8}$ of massive galaxies, more than $\sim 10\%$ of FRBs would have to be sourced in galaxies of $M_{\rm halo}=10^{14}\,M_\odot$ for this correlation to be comparable to the one due to the WHIM.

\section{Method}
\label{sec:Method}

We will now outline the method to cross correlate the two observables that we have described: DMs and $y$ maps.

The angular resolution of current CMB experiments is on the scale of $1-10$ arcmin. For instance, the Planck experiment has a full-width half-maximum resolution of $\theta_{\rm FWHM}=5$ arcmin in the 217 GHz channel~\cite{Lamarre:2003zh}, whereas both ACT and SPT have $\theta_{\rm FWHM}\approx 1$ arcmin~\cite{Swetz:2010fy,Austermann:2012ga}, which is comparable to what is expected of the CMB S4 experiment~\cite{Abazajian:2016yjj}.
On the FRB side, upcoming observatories, such as CHIME~\cite{Bandura:2014gwa,ChimeFRB}, HIRAX~\cite{Newburgh:2016mwi}, or HERA~\cite{DeBoer:2016tnn}, will also possess angular resolutions of order arcminutes.
For instance, it is expected that CHIME will localize FRBs to within~\cite{Masui2017}
\be
\Delta\theta \approx 1 {\rm \ arcmin} \times \left( \dfrac{\rm SNR}{10}\right)^{-1} \times \left( \dfrac{\nu}{600\ \rm MHz}\right)^{-1},
\ee
where we have chosen the central frequency to be $\nu=600$ MHz, and a signal-to-noise ratio SNR $=10$ as the threshold to claim a detection. This clearly shows that every FRB will be localized to an area comparable to, or smaller than a CMB pixel.

One can use the full resolution of the FRBs by individually correlating the DM and $y$ parameter for each detection\footnote{Note that, with arcmin resolution, it would take about $10^8$ FRBs to have a full-sky map of the integrated DM, whereas we only expect to have $10^4$ detections per year. The event-by-event correlations that we propose can extract the maximum amount of information from the DM detections.}. 
Given that the DM and the $y_{\rm whim}$ component are sourced by the same intergalactic gas, we would expect a positive correlation between them.
In order to demonstrate an implementation of this method, we will
perform a Monte Carlo simulation, in which we model the effects that underlie in the correlation between $y$ and DM maps, and we will forecast how many measured DMs will be required to observe the cross correlation between DM and $y$ in the WHIM.

\subsection{Monte Carlo Setup}

We assume that FRB sources are distributed as a smooth function of redshift. From the observed DMs one can infer a redshift, assuming all the DM is due to the IGM. That distribution is consistent with a constant comoving number density with a cutoff in luminosity~\cite{Munoz:2016tmg},
\be
\dfrac{dN_{\rm FRB}}{dz} = \mathcal N \dfrac{c\, \chi^2(z)}{(1+z)H(z)} e^{-d_L^2(z)/d_L^2(z_{\rm cut})},
\label{eq:Nz}
\ee
where $\chi$ is the radial comoving distance, $d_L$ is the luminosity distance, and there is an extra factor of $(1+z)$ in the denominator to account for the redshift in the temporal rate of FRBs. A cutoff of $z_{\rm cut}=0.5$ is consistent with current FRB data (see also Ref.~\cite{Luo:2018tiy}).
This function has to be interpreted as an estimate, and future observations will enable us to pinpoint exactly the shape of this function.
We choose the normalization constant $\cal N$ such that $\int (dN_{\rm FRB}/dz) dz =  N_{\rm FRB}$, where we vary the number $N_{\rm FRB}$ of detected FRBs.

Working with real data, one would have to take into account the varying angular localizations of different FRBs, depending on their signal-to-noise ratios. 
Additionally, FRB localization areas could intersect multiple CMB pixels, or more than one FRB can be coincident with the same CMB pixel, in which case an appropriate average over the signal in each pixel should be taken. 
Here we will assume that each FRB is assigned a unique value of $y$, after all averaging is done.

In our Monte Carlo simulation  we generate FRBs with redshifts extracted from the PDF in Eq.~\eqref{eq:Nz}.
Each FRB is then given an IGM contribution to their DM through Eq.~\eqref{eq:DMz}, and a random host contribution with a flat PDF,
\be
P({\rm DM_{host}}) = \Theta(2\, \overline{\rm DM}_{\rm host}(z)-{\rm DM_{host}}),
\ee
running between 0 and twice the average value, given by Eq.~\eqref{eq:DMhost}.
Additionally, we assign each FRB a $y_{\rm whim}$ given by Eq.~\eqref{eq:ywhim} with Model I integrated up to its origin redshift $z$, to which we add a noise component extracted from a Gaussian distribution with width $y_{\rm noise} = 10^{-6}$, as observed in Planck maps~\cite{Aghanim:2015eva}.

\subsection{Simulation Output}

\begin{figure}[hbtp!]
	\includegraphics[width=0.49\textwidth]{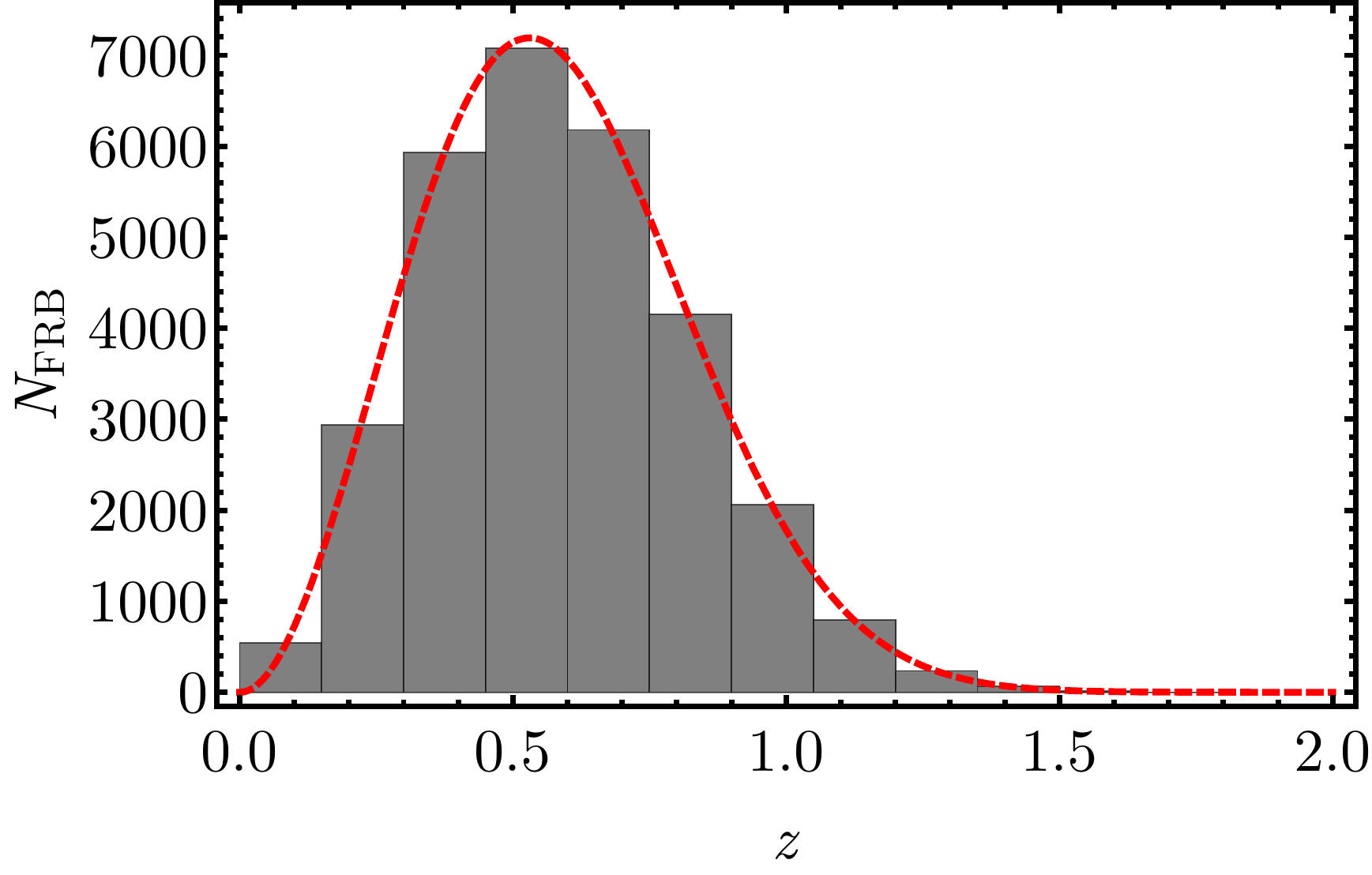}
	\caption{Redshift distribution of our mock FRB population. The gray bins represent the histogram of a Monte Carlo run with $30,000$ FRBs, and the dashed-red line is the input PDF. 
	}
	\label{fig:dNz}
\end{figure}

\begin{figure}[hbtp!]
	\includegraphics[width=0.44\textwidth]{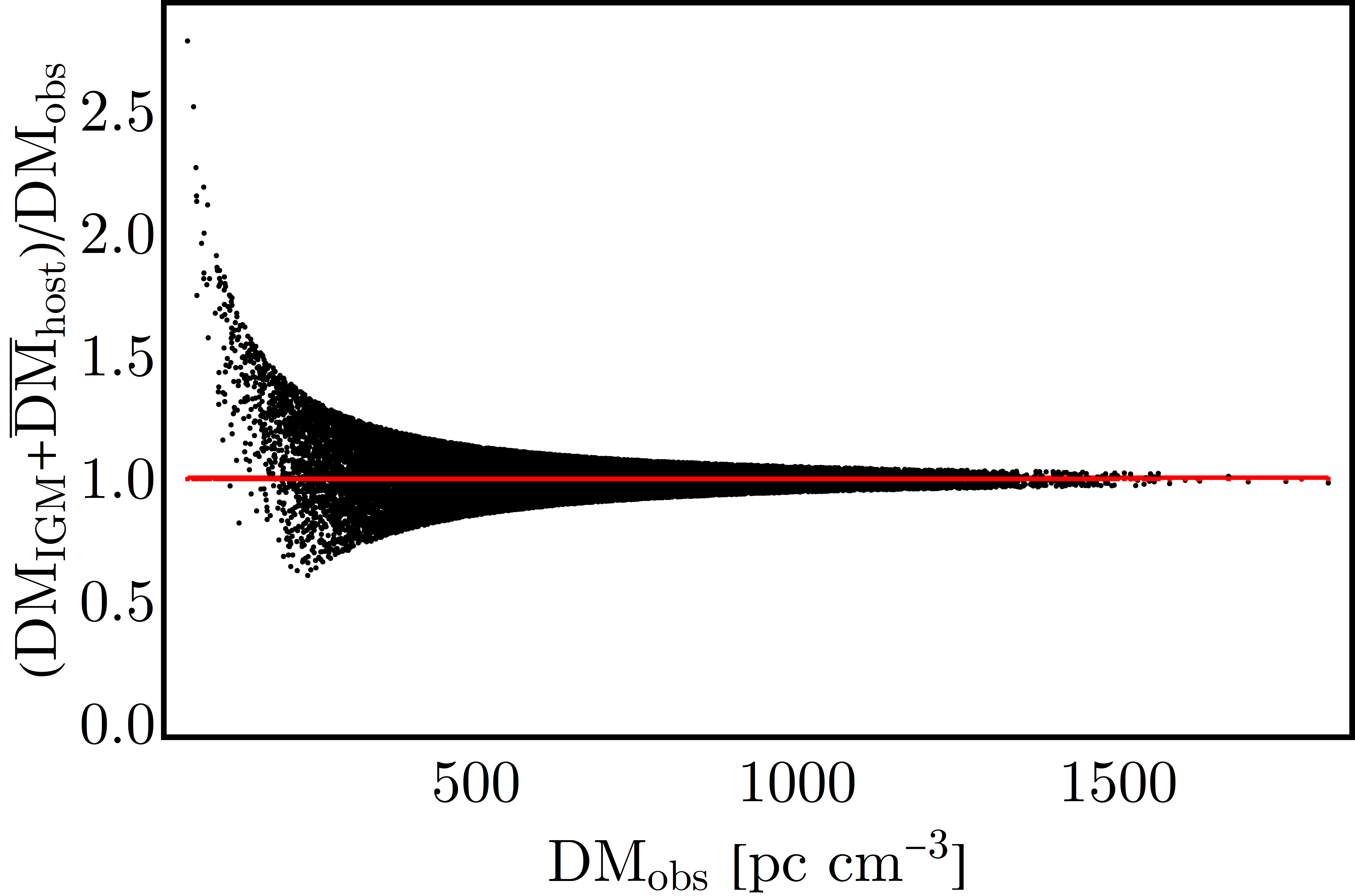}
	\caption{Deviation from the expected DM for each FRB, obtained by adding the DM$_{\rm IGM}$ component due to the IGM to the average due to the host, $\rm \overline{ DM}_{\rm host}$$(z)$, versus the observed DM. The red line follows unity.
	}
	\label{fig:DMz}
\end{figure}

We run a Monte Carlo simulation with $N_{\rm FRB}=30,000$ mock FRBs, corresponding to a few years of CHIME or HIRAX data.
Fig.~\ref{fig:dNz} shows the input PDF for the FRB redshift distribution, as well as the result of this first simulation. 
In Fig.~\ref{fig:DMz} we study the host contribution to the DM, by plotting the ratio of the expected DM (given by the sum of the IGM and average-host contributions) to the observed DM$_{\rm obs}$. 
This figure showcases how low DM$_{\rm obs}$, corresponding to low redshifts, can obtain a significant contribution from the FRB host, whereas large-$z$ FRBs acquire most of their DM from the IGM.

\begin{figure}[hbtp!]
	\includegraphics[width=0.49\textwidth]{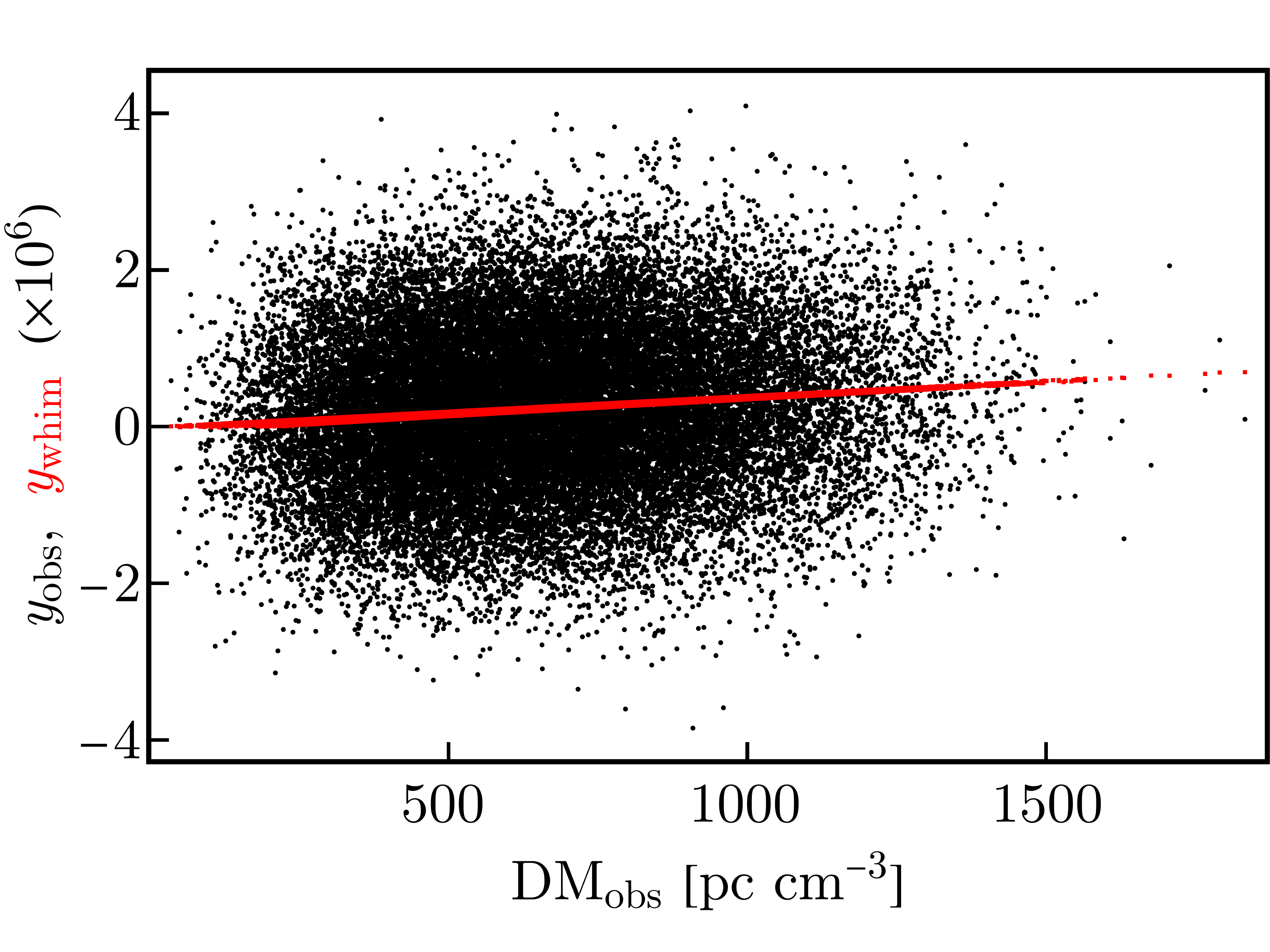}
	\caption{DM-$y$ correlation for our mock data. We show in black the observed Compton $y$ parameter, including noise, versus the DM for each of our 30,000 simulated FRBs.
	In red we show the WHIM component of the $y$ parameter, clearly subdominant, albeit correlated with DM.
	}
	\label{fig:yvsDM}
\end{figure}

We show the observed value of $y$, as well as the underlying (and not directly observable) $y_{\rm whim}$ contribution in Fig.~\ref{fig:yvsDM}, where the correlation between these two variables is clear, albeit hidden behind the noise. 
We will now show how to extract the DM-$y$ correlation from the mock data in the presence of a significant $y$ noise.

\subsection{Cross Correlation}

Given a list of values of DM and $y$, associated with each FRB, we find the correlation coefficient between these two variables simply as
\be
r ({\rm DM},y) = \VEV{{\rm DM}\, y} - \VEV{\rm DM}\VEV{y}.
\label{eq:ryDM}
\ee
Notice that, in addition to $\VEV{\rm DM}$ not being zero (as it is a positive-definite quantity), $\VEV{y}$ need not be zero either, as the contribution from collapsed structures skews $y$ positively. We will revisit this point.
We can obtain an order-of-magnitude estimate of this quantity by remembering that for a typical FRB (from $z=0.5$) DM$_{\rm IGM} \sim10^3$ pc cm$^{-3}$, whereas for Model I of the WHIM, $y_{\rm whim} \sim 10^{-7}$.
We estimate, then, $r ({\rm DM},y) \sim 10^{-4}$ pc cm$^{-3}$.
We calculate the correlation coefficient $r($DM,$y$) for our simulated data with Eq.~\eqref{eq:ryDM}, and find it to be 
\be
r({\rm DM},y) = 2 \times 10^{-4}\,\rm pc \, cm^{-3},
\ee
in line with our expectations, showing the significant DM-$y$ correlation in the simulated data.

\begin{figure}[hbtp!]
	\includegraphics[width=0.49\textwidth]{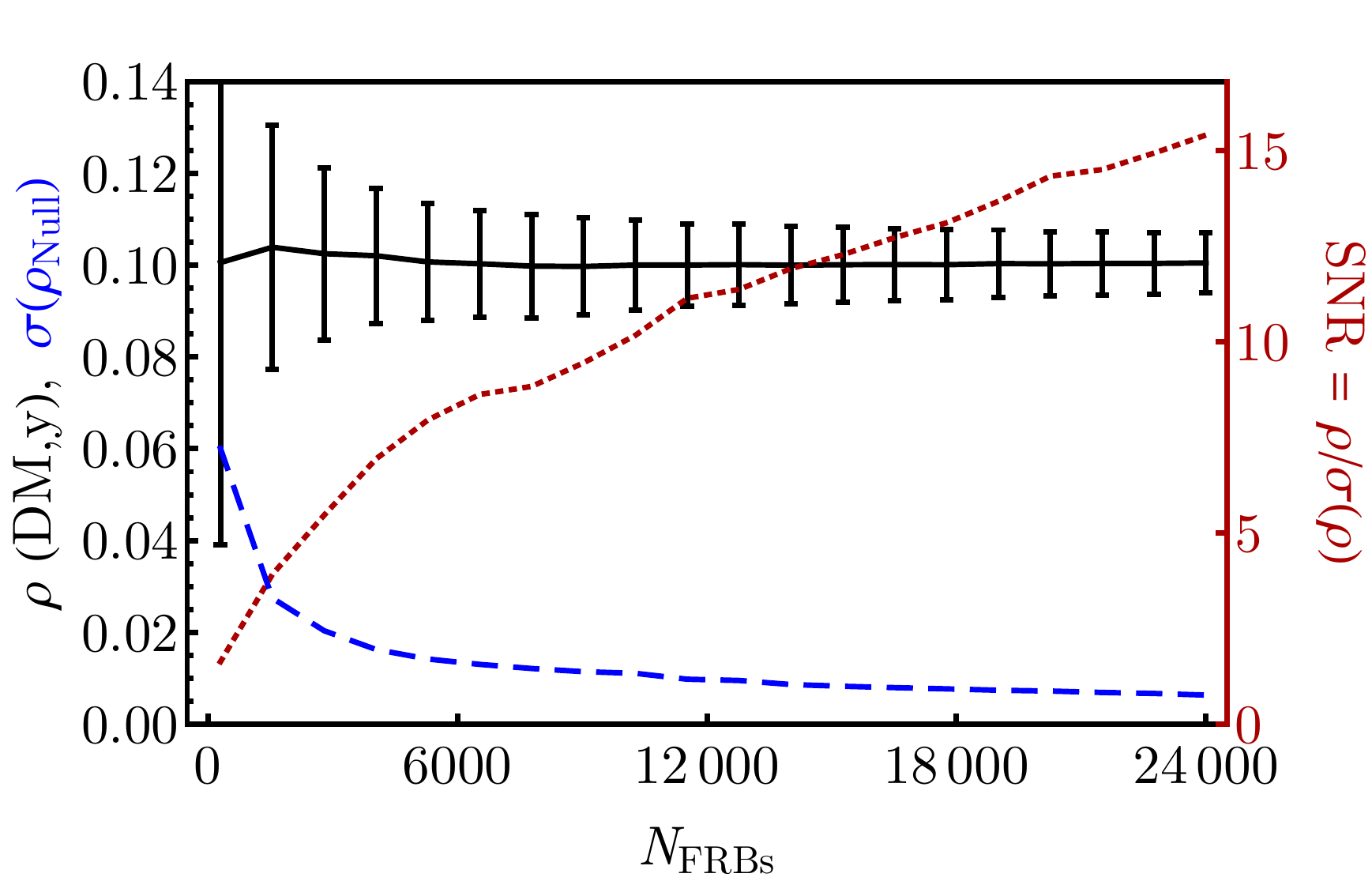}
	\caption{
The black line shows the cross-correlation coefficient $\rho({\rm DM}, y)$ between DM and $y$ for a simulation as specified in Sec.~\ref{sec:Method}, versus the number $N_{\rm FRB}$ of observed FRBs. The error bars are obtained via bootstrapping of 30,000 FRBs, and the dashed blue line represents the the standard deviation $\sigma(\rho_{\rm Null})$ of the $\rho_{\rm Null}$ of 300 null tests. 
	Finally, the dotted-red line shows the ratio between $\rho$ and its error, as a measurement of the signal-to-noise ratio (SNR).
	}
	\label{fig:rxy}
\end{figure}

We define the cross-correlation parameter 
\be
\rho({\rm DM},y) = \dfrac{r({\rm DM},y)}{\sigma_{\rm DM} \sigma_y}
\label{eq:rhoyDM}
\ee
in the usual way, where $\sigma_X$ is the standard deviation of $X$, directly calculated from the mock data.
For our simulation we have found a (dimensionless) cross correlation parameter $\rho($DM,$y$)$=0.1$.
This exemplifies the large correlation between the two variables, DM and $y$, due to the WHIM.
We show this cross-correlation coefficient $\rho$ in Fig.~\ref{fig:rxy} as a function of the number $N_{\rm FRB}$ of observed FRBs. 
We find the error in $\rho$ by bootstrapping, for which we pick $N_{\rm FRB}$ from the initial $30,000$ FRBs with replacement. 
These are the error bars shown in Fig.~\ref{fig:rxy}, which approximately decrease as $1/\sqrt{N_{\rm FRB}}$ with the number $N_{\rm FRBs}$ of FRBs, as expected from Poisson noise.
Moreover, in order to ensure we are observing a real correlation, we have performed a null test, where we generate DMs the usual way, but do not inject the $y_{\rm whim}$ signal. 
We show in Fig.~\ref{fig:rxy} the standard deviation of 300 of these null tests for comparison. The error from boostrapping and the standard deviation of this null test are very similar, as expected given the small contribution of $y_{\rm whim}$ to the overall $y$ noise.
Note that here we are not injecting a correlated tSZ signal from clusters, given by the $\left. y_{\rm cl}\right|_{\rm FRB}$ from Eq.~\eqref{eq:ycluster}, so in the null tests the correlation is expected to be zero.
From Fig.~\ref{fig:rxy} it can be seen that for $N_{\rm FRB}\gtrsim 5,000$ the ratio of $\rho$ to its noise is $\gtrsim5$, making this a prospective detection, and for $N_{\rm FRB}\gtrsim 25,000$ the SNR would be $\gtrsim 15$.
Note that here we are using Model I of the WHIM, which has $f_{\rm whim}=1$, and these results should be rescaled by $f_{\rm whim}$ for other fractions, as discussed below.

\subsection{Tests}

Next we test how the predicted signal changes under different assumptions about both the properties of the FRBs and the Gaussianity of the tSZ map.

\subsubsection{FRB Distribution}

In the previous figures we assumed a population of FRBs with $z_{\rm cut}=0.5$, as in Ref.~\cite{Munoz:2016tmg}. This yielded $r({\rm DM},y) = 2 \times 10^{-4}$ pc cm$^{-3}$, and given the uncertainties of the instruments outlined above, a cross correlation parameter of $\rho({\rm DM},y) = 0.1$.
Newly observed FRB show large DMs (see, for instance, Ref.~\cite{Bhandari:2017qrj}), so we will 
entertain the possibility of a higher-$z$ FRB population, with $z_{\rm cut}=1$ and the same functional form as Eq.~\eqref{eq:Nz}.
In this case, using Model I we find a larger cross correlation of $r({\rm DM},y) = 6 \times 10^{-4}$ pc cm$^{-3}$, yielding $\rho({\rm DM},y) = 0.18$. This would greatly enhance the detectability of this signal, although it would require a factor of $\sim 5$ better flux limit in the radio. 
Observatories with greater sensitivities to fainter FRBs have, thus, better prospects to detect these cross correlations.

Additionally, we briefly consider the impact of the unknown host contribution to the DM to the cross correlation.
In the original simulation we chose a non-Gaussian PDF for the host contribution, in Eq.~\eqref{eq:DMhost}, flat in the 0 to 2$\,\overline{\rm DM}_{\rm host}(z)$ range.
We have ran a simulation without this component and found a negligible difference in both $r({\rm DM},y)$ and $\rho({\rm DM},y)$.
This is to be expected, since the variation of the DM$_{\rm IGM}$ between different FRBs dominates over the host component of $\sigma_{\rm DM}$, and the host contribution does not correlate with $y$ in any of our models.
Nonetheless, this confirms that our assumptions about $\overline{\rm DM}_{\rm host}(z)$ do not affect the resulting DM-$y$ correlation.

Finally, while our method does not explicitly account for clustering along the line of sight (see, e.g. Refs.~\cite{Zhang:2007psa,McQuinn:2013tmc}), the one-to-one mapping between DMs and $y$ in our MonteCarlo simulations implicitly includes this effect, as larger DMs (from higher-$z$ FRBs) are accompanied by larger values of $y_{\rm whim}$.
This, however, assumes that the IGM and the WHIM have similar biases, which given the uncertainties in the clustering properties of the WHIM is to be confirmed.

\subsubsection{Non-Gaussianity of the tSZ Map}

So far we have ignored the non-Gaussianity intrinsic to the tSZ map, and parametrized the PDF for the Compton $y$ parameter as a Gaussian centered around zero, with width $y_{\rm noise} = 10^{-6}$. Nonetheless, it is well-known that the signal from collapsed structures (such as clusters) induces non-Gaussianities in the $y$ map, skewing it positively~\cite{Zhang:2007psa}. 
In this subsection we will explore the effect of this skewness in our results.

\begin{figure}[hbtp!]
	\includegraphics[width=0.46\textwidth]{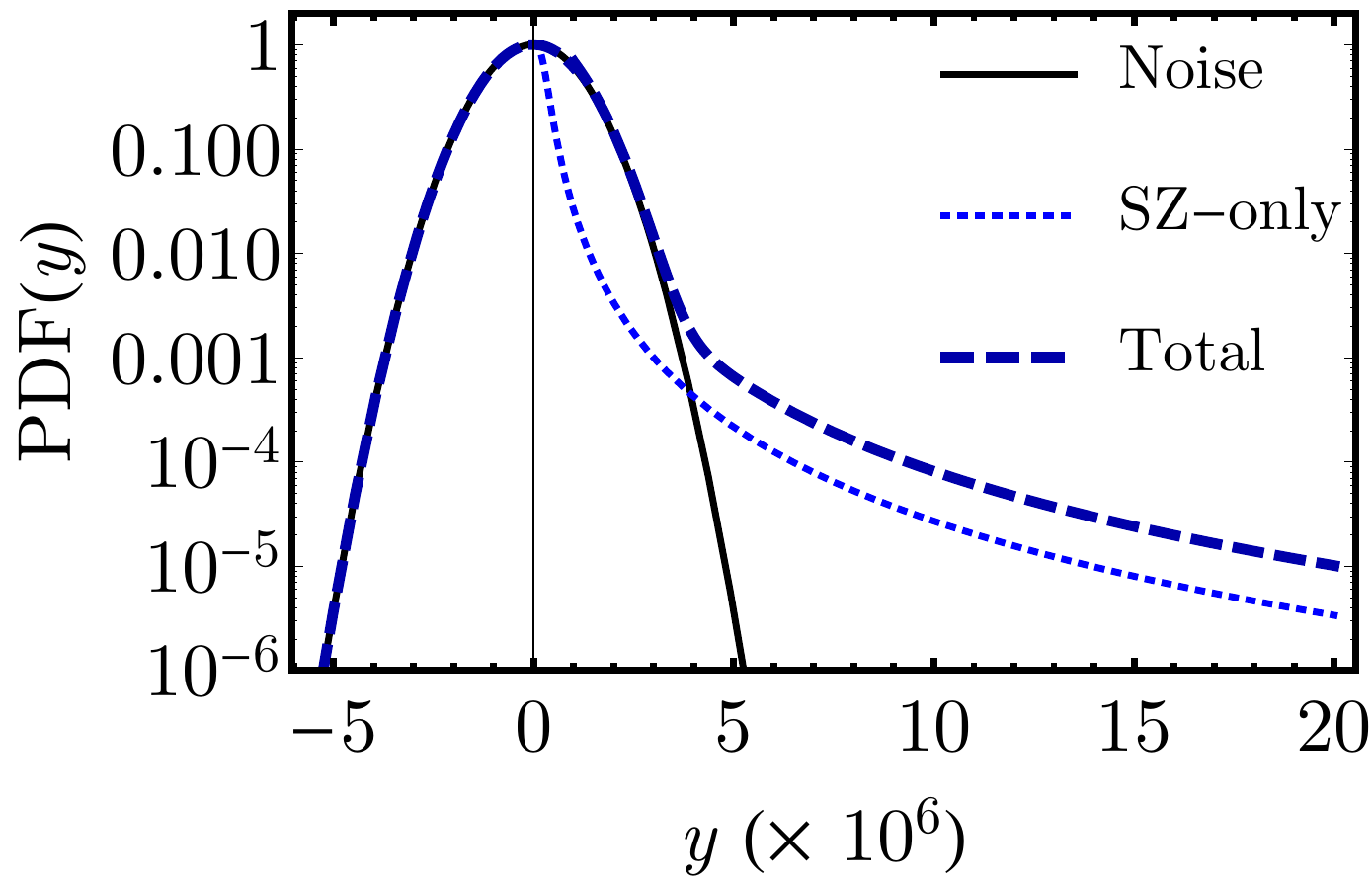}
	\caption{Different PDFs for the Compton $y$ parameter. We show the noise-only case with $y_{\rm noise} = 10^{-6}$ as the solid line, the SZ-only case from Eq.~\eqref{eq:PDFyNG} as the thin-dotted line, and the total PDF with $A_{\rm NG}=3$ (described in Eq.~\eqref{eq:PDFytot}) as the thick-dashed line.
	}
	\label{fig:PDFy}
\end{figure}

We will not attempt to model the non-Gaussianity of the $y$ map from first principles, and instead just fit the result from the Planck satellite~\cite{Aghanim:2015eva}, 
where the ``noise-only" PDF of $y$ is, to a good approximation, Gaussian, and given by
\be
P_{\rm noise}(y) = e^{-y^2/(2 y_{\rm noise}^2)},
\label{eq:PDFynoise}
\ee
whereas the ``SZ-only" PDF provides a non-Gaussian tail that can be fit by
\be
P_{\rm SZ-only}(y) = \dfrac{1}{1+\left( y/\tilde y\right)^\alpha}
\Theta(y-y_{\rm noise}), 
\label{eq:PDFyNG}
\ee
where $\Theta$ is the Heaviside Theta function, chosen to avoid the contribution from the power-law tail at small or negative $y$.
We find that $\tilde y \approx 0.3$ and $\alpha\approx3$ can fit the data from Ref.~\cite{Aghanim:2015eva}.
We normalize the total PDF with a relative amplitude $A_{\rm NG}$, which controls the size of the non-Gaussianities, to find
\be
P_{\rm total} (y) = P_{\rm noise}(y) + A_{\rm NG} P_{\rm SZ-only}(y).
\label{eq:PDFytot}
\ee
We show the three PDFs for $y$ in Fig.~\ref{fig:PDFy}, where we find that $A_{\rm NG}=3$ provides a good fit for the total PDF from Ref.~\cite{Aghanim:2015eva}.
We emphasize that our fit should not be taken as a substitute for the real $y$ PDF, but instead as a computationally simple recipe to generate non-Gaussian $y$-parameter realizations.

\begin{figure}[hbtp!]
	\includegraphics[width=0.49\textwidth]{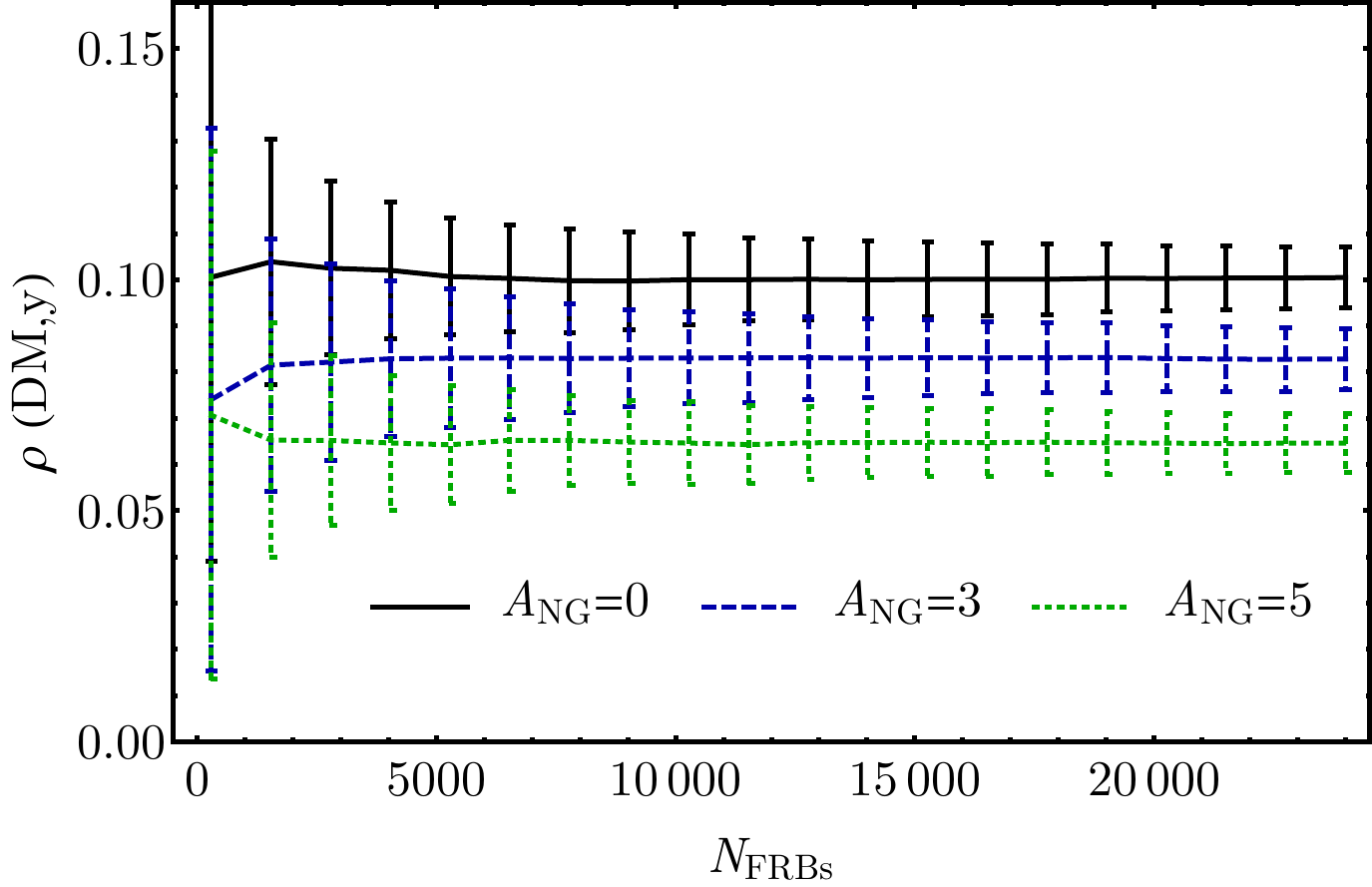}
	\caption{Same as Fig.~\ref{fig:rxy}, albeit considering non-Gaussianities in the PDF of $y$, as given by Eq.~(\ref{eq:PDFytot}).
	}
	\label{fig:rxy_NG}
\end{figure}

We perform the same analysis as before, albeit generating noise in $y$ with Eq.~\eqref{eq:PDFytot}, and we show the  cross-correlation parameter in Fig.~\ref{fig:rxy_NG} for three cases, $A_{\rm NG}=0$, 3, and 5.
We see that a more non-Gaussian $y$ mildly lowers $\rho({\rm DM},y)$, given the increase in the noise. 
Note, however, that in the previous calculations we have already accounted for  broadening in the $y$ PDF, by setting $y_{\rm noise} = 10^{-6}$, roughly 40\% larger than inferred by the ``Noise-only" curve in Ref.~\cite{Aghanim:2015eva}.
As for the correlation coefficients, we find $r({\rm DM},y)=3\times 10^{-4}$ pc cm$^{-3}$ for both $A_{\rm NG}=3$, and 5, slightly larger than in the $A_{\rm NG}=0$ case due to the larger typical values of $y$.
Thus, we conclude that the non-Gaussianities in the $y$ map do not qualitatively change the results that we present, although they can modify specific values of the DM-$y$ correlation within tens of percent.

\section{Applications}
\label{sec:App}

In the previous section we outlined how to effectively cross correlate DMs (obtained from FRBs) and tSZ $y$ maps (from CMB experiments).
We will now study three applications of this method.

\subsection{Confirming the Presence of the WHIM}

The first, and most straightforward, application of DM-tSZ cross correlations is to confirm the WHIM detection from Refs.~\cite{Tanimura:2017ixt,deGraaff:2017byg} (see also Ref.~\cite{Nicastro:2018eam}).
By assuming a homogeneous temperature in the filament, and some value of the central overdensity, it was estimated in Ref.~\cite{deGraaff:2017byg} that the filamentary WHIM is characterized by the product
\be
f_{\rm whim} T_{\rm whim} \approx 0.3 \times 10^6 \rm \, K.
\label{eq:tSZwhimdet}
\ee

We found in the previous section that we expect $\rho({\rm DM},y)=0.1$ for Model I, which has a fraction of baryons in the WHIM of $f_{\rm whim}=1$, and temperature today of $T_e^{(0)}=100$ eV.
More realistically, we don't expect all baryons to be in the WHIM, so  we divide the IGM into a cold component, which contributes to the dispersion measure but does not produce a tSZ effect; and a WHIM component, which contributes to both.
The tSZ effect of the WHIM component, given by Eq.~\eqref{eq:ywhim}, will be sensitive to the product of $f_{\rm whim}$ and its temperature $T_{\rm whim}$. By rescaling the results that we found in Sec.~\ref{sec:Method} to the values of Eq.~\eqref{eq:tSZwhimdet} we find
\be
r({\rm DM},y) = 5 \times 10^{-5}\,{\rm pc\, cm^3} \left( \dfrac{T_{\rm whim}}{10^6\,\rm K}\right)\left( \dfrac{f_{\rm whim}}{0.3}\right),
\ee
assuming constant $T_{\rm whim}$.
This value of $r({\rm DM},y)$ translates into a cross-correlation parameter
\be
\rho({\rm DM},y) = 0.026 \times \left( \dfrac{T_{\rm whim}}{10^6\,\rm K}\right)\left( \dfrac{f_{\rm whim}}{0.3}\right),
\ee
for our noise parameters, which is observable at an SNR of $\gtrsim 5$ for 30,000 FRBs, and SNR $\gtrsim 10$ for a higher-$z$ FRB population, with $z_{\rm cut}=1$.
It is, then, within the reach of upcoming FRB observations, and current CMB maps, to confirm the WHIM detection of Refs.~\cite{Tanimura:2017ixt,deGraaff:2017byg}.

\subsection{An Intergalactic Thermostat}

We now proceed beyond the preliminary question of whether the signal is detectable, and study how to find the origin of a prospective DM-$y$ cross-correlation detection.

In all calculations above we assumed that the electron temperature $T_e(z)$ scales as $(1+z)^{-1}$, as in Model I of Section~\ref{sec:tSZ}.
Nonetheless, if hot intergalactic gas (or any other source) causes both DM and $y$, we can use Eqs.~(\ref{eq:DMz},\ref{eq:ywhim}) to find the average temperature of electrons as a function of redshift:
\be
T_e(z) = \dfrac{m_e c^2}{(1+z) k_B \sigma_T}\, \left.\dfrac{dy}{d(\rm DM)}\right|_z.
\ee
Therefore, given sufficiently precise measurements of DM and $y$, one can find the electron temperature as a function of redshift.
From this relation it is easy to see that for Model I, with $T_e(z)\propto(1+z)^{-1}$, $y_{\rm whim} \propto \rm DM_{\rm IGM}$. 
The  shock heating of the WHIM, sourced by structure formation, is only predicted to significantly heat the WHIM at low redshifts ($z<3$)~\cite{Cen:1998hc,Haider:2015caa}.
Therefore, inspired by the simulations in Ref.~\cite{Cen:1998hc} we surmise a Model II of the WHIM, for which
\be
T_e(z) = T_e^{(0)} e^{-\alpha z},
\ee
with fiducial values of $T_e^{(0)}=10^6$ K and $\alpha=3/2$.
We keep $f_{\rm whim}$ fixed to its $z=0$ value, which we set $f_{\rm whim}=0.3$, as suggested by Refs.~\cite{Tanimura:2017ixt,deGraaff:2017byg}; and absorb its redshift dependence into $T_e(z)$, which follows the density-averaged temperature of Ref.~\cite{Cen:1998hc}.
In that case, $y$ would increase with redshift slower than in Model I. By substituting $T_e(z)$ in Eq.~\eqref{eq:ywhim} we find
\be
y_{\rm whim} (z) = f_{\rm whim} \dfrac{k_B T_{\rm gas}^{(0)} n_e^{(0)}\,\sigma_T}{m_e c} \int_0^z \dfrac{dz' (1+z')^2 e^{-\alpha z}}{H(z')}.
\label{eq:ywhimCen}
\ee

\begin{figure}[hbtp!]
	\includegraphics[width=0.49\textwidth]{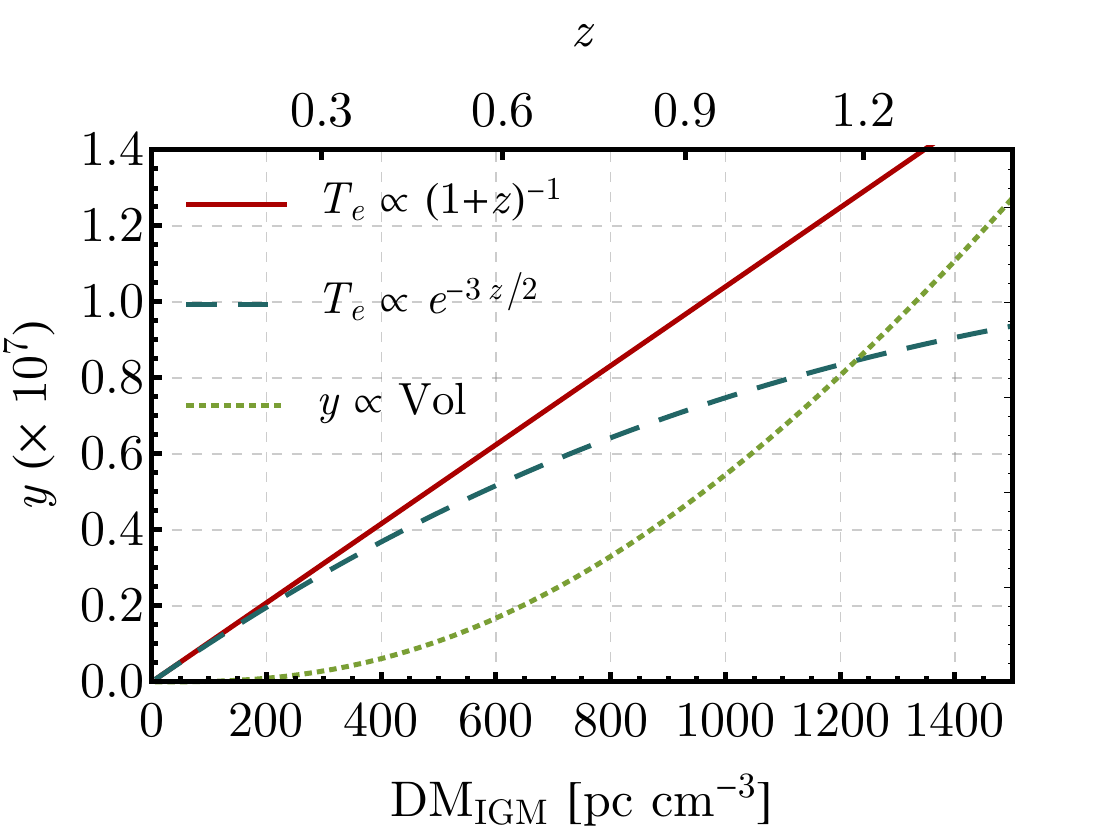}	
	\caption{ Behavior of the Compton $y$ parameter as a function of DM$_{\rm IGM}$ (including whim and non-whim components; in pc cm$^{-3}$) for three models. In solid-red line we show $T_e(z)\propto (1+z)^{-1}$, in dashed-blue line Model II, as in Eq.~\eqref{eq:ywhimCen}, and in long-dashed-black line $y$ proportional to comoving volume, as a proxy for the signal sourced by clusters.
	}
	\label{fig:yvsDMsources}
\end{figure}

We show the value of $y_{\rm whim}$ for Model II in Fig.~\ref{fig:yvsDMsources}, as a function of DM$_{\rm IGM}$ (which includes WHIM as well as non-WHIM contributions).
We also plot a rescaled version of Model I, with $T_e = 10^6 \,{\rm K} \times (1+z)^{-1}$ and $f_{\rm whim}=0.3$, which clearly shows a larger tSZ signal at higher DMs (and therefore larger redshifts).
Additionally, in Fig.~\ref{fig:yvsDMsources} a putative $y$ contribution from clusters is shown, where we assumed a constant number density of sources, so that $y$ is proportional to comoving volume.
We normalized this contribution to obtain $\left. y_{\rm cl}\right|_{\rm FRB}(z=0.5)=10^{-8}$.

\begin{figure*}[hbtp!]
	\includegraphics[width=0.7\textwidth]{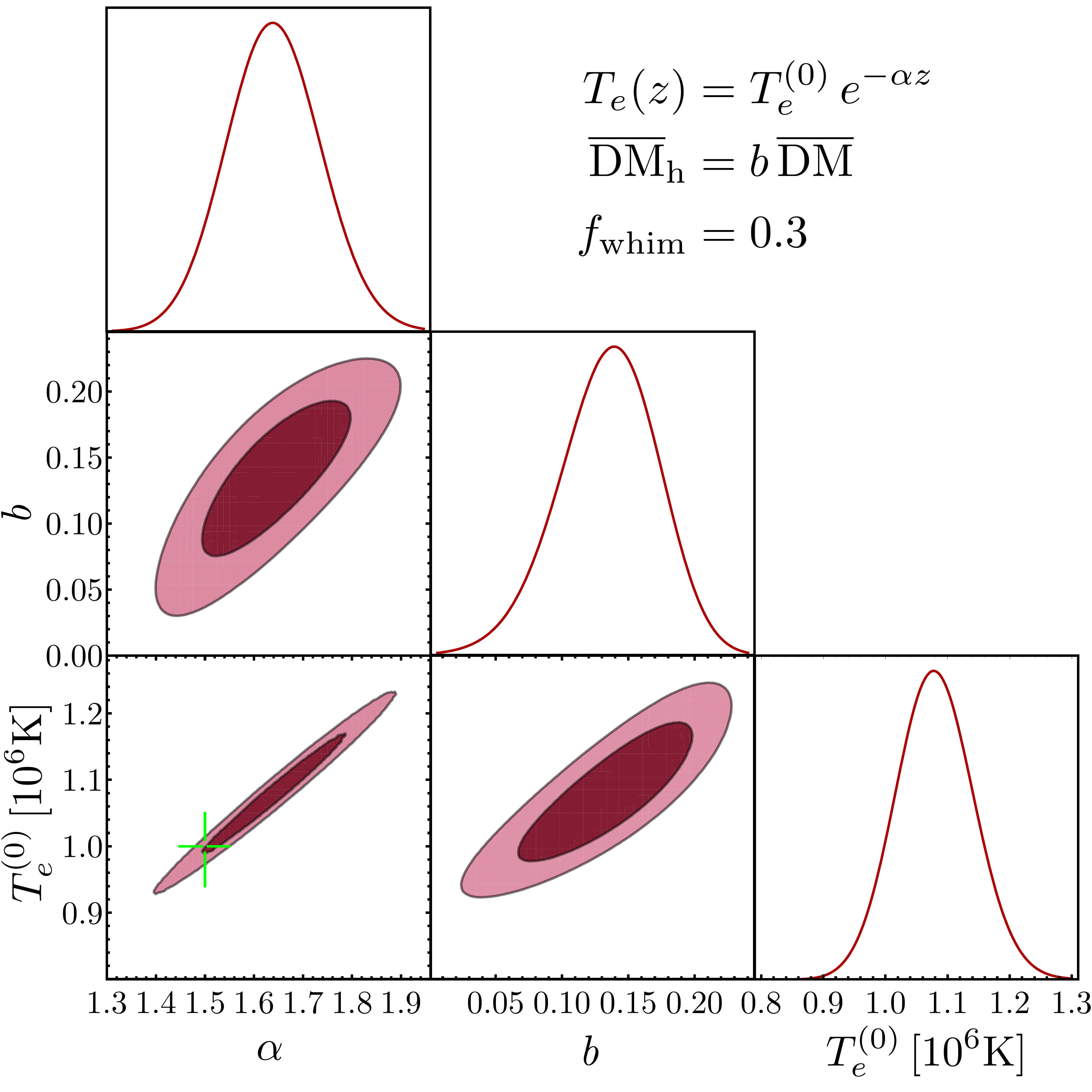}	
	\caption{Confidence ellipses (68\% C.L.~in dark red and 95\% C.L.~in light red) for the fit of Eq.~\eqref{eq:yfitDM}.
		Here we have assumed Model II of the WHIM, with $T_e(z) = T_e^{(0)} e^{-\alpha z}$, and the host contribution is parametrized through $b$ as in Eq.~\eqref{eq:bdef}.
		We have assummed $N_{\rm FRB}=30,000$ observed DMs, and $y_{\rm noise}= 10^{-7}$.
		The green cross indicates our input values of $T_e^{(0)}=10^6$ K and $\alpha=3/2$, whereas $b$ is set by the physics of the FRB host galaxy.
		We also plot the likelihood for each variable, once marginalized over the rest.
	}
	\label{fig:ellipses}
\end{figure*}

\subsubsection*{Parameter Estimation with $y$-DM correlation}

Let us now explore further the idea of using DM-$y$ cross correlations as an intergalactic thermostat, by finding what constraints can be achieved by future data on the fiducial parameters of our Model II.

It is very likely that we will not be able to assign a redshift to every detected FRB.
In that case, the observed DM can act as a proxy of redshift, as DM$_{\rm IGM}$ and $z$ can be linearly related via Eq.~\eqref{eq:DMzlinear}.
This allows us to find the $y_{\rm whim}$ component, instead of as a function of $z$, in terms of the observed DM.
Therefore, we posit a model given by Eq.~\eqref{eq:ywhimCen}, which we can analytically integrate for ${\rm DM_{IGM}}(z) = \overline{\rm DM}\,z$ to find
\be
y^{\rm model}({\rm DM};T_e^{(0)},\alpha,b) = T_e^{(0)} C g({\rm DM},\alpha),
\label{eq:yfitDM}
\ee
with $C = f_{\rm whim} \overline{\rm DM} \, \sigma_T k_B/(m_e\,c^2)\approx 1.07\times 10^{-13}\,K^{-1}$ for our fiducial parameters,  and we have defined the function
\be
g({\rm DM},\alpha) = \dfrac{1+\alpha-e^{-\alpha \tilde z} (1 + \alpha + \alpha \tilde z)}{\alpha^2},
\ee
where the linearized redshift $\tilde{z}$ is obtained by reversing Eq.~\eqref{eq:DMzlinear}, and approximately subtracting a host contribution, to be
\be
\tilde z = \dfrac{{\rm DM}}{\overline{\rm DM}} - \dfrac{b}{1+\dfrac{{\rm DM}}{\overline{\rm DM}}},
\label{eq:ztilde}
\ee
where the parameter $b$ expresses the average host contribution to the DM through
\be
b = \dfrac{\overline{\rm DM}_h}{\overline{\rm DM}},
\label{eq:bdef}
\ee
where we remind the reader that $\overline{\rm DM} = 1025$ pc cm$^{-3}$, and that in all of our inputs we assume no host evolution, and thus $\left. \overline{\rm DM}\right._{\rm host} = \overline{\rm DM}_h (1+z)^{-1}$.
It is, then, the three parameters $T_e^{(0)}$, $\alpha$, and $b$ that we will constrain with mock data.

To obtain the mock data we have ran a simulation with $N_{\rm FRB}=30,000$ FRBs, as in Sec.~\ref{sec:Method}, and stored the values of $y$ and DM observed, both of which include the WHIM and an unrelated component (noise in the case of $y$ and non-WHIM IGM for DM).
Given the challenge of constraining the thermal state of the IGM, we will take a higher-redshift FRB population, setting $z_{\rm cut}=1$ in Eq.~\eqref{eq:Nz}, and will assume $y_{\rm noise}=10^{-7}$, as an optimistic estimate of the capabilities of future CMB experiments, including perhaps the masking of a significant fraction of tSZ-emitting haloes.

Given the observed \{DM$_i^{\rm obs}$, $y_i^{\rm obs}$\} pairs, we compute the likelihood at each value in parameter space, assuming Gaussianity, through the chi-squared statistic,
\be
\chi^2(T_e^{(0)},\alpha,b) = \sum_{i} \left[ \dfrac{y^{\rm obs}_i - y^{\rm model}({\rm DM_i^{\rm obs}};T_e^{(0)},\alpha,b) }{\sigma_i}\right]^2,
\ee
where we further simplify $\sigma_i \equiv \sigma(y) = y_{\rm noise}$, given that the DM errors are significantly smaller than those of $y$. Then the likelihood is simply computed as $\mathcal L = {\rm exp}(-\chi^2/2)$.

We show the confidence ellipses, and likelihoods, for the three parameters in Fig.~\ref{fig:ellipses}, where our inputs are $T_e^{(0)}=10^6$ K and $\alpha = 3/2$, with $f_{\rm whim}=0.3$.
From our model of the host contribution we expect $\overline{\rm DM}_h =$ 100 pc cm$^{-1}$ (corresponding to $b \approx 0.1$), although its value is mildly redshift dependent, even after applying Eq.~\eqref{eq:ztilde}, since DM $\propto z$ only approximately.
From our simulated data we obtain a measurement of
\begin{align} 
T_e^{(0)} &= 1.08^{+0.13}_{-0.12} \, \times 10^6 \,\rm K \quad  and \nonumber \\ 
\alpha &= 1.63^{+0.21}_{-0.19},
\end{align}
both at 95\% C.~L.
This agrees with our inputs within ten percent, showing the promise of DM-tSZ cross correlations to measure the thermal state of the IGM.
Note that, as was the case for $r({\rm DM},y)$, we are only sensitive to the product $f_{\rm whim} T_e^{(0)}$.
Given that we have fixed  $f_{\rm whim} = 0.3$ here, the exponent $\alpha$ determines the decay of the density-averaged $T_e$.

Moreover, our analysis can potentially shed light on the contribution of the FRB host to the DM, through the parameter $b$. 
We find a best-fit value of $b=0.14^{+0.07}_{-0.08}$, corresponding to a local averaged host contribution
\be
\overline{\rm DM}_h = 141^{+75}_{-85} \, \rm pc\,cm^{-3}
\ee
at 95\% C.L., firmly 2-$\sigma$ away from zero.
This would be a unique way to disentangle the host contribution to the DM from that of the IGM, breaking degeneracies in FRB studies for cosmology~\cite{Walters:2017afr}.
Notice that due to the approximate linear nature of the DM-$z$ relation, and the high correlation of  both $\alpha$ and $T_e^{(0)}$ with $b$, the best-fit value of both parameters is roughly 1-$\sigma$ biased with respect to our inputs.

Additionally, we have checked that with $y_{\rm noise}=3\times 10^{-7}$, only a factor of $\sim$2$-$3 better than current noise levels, the constraints on the parameters would still be competitive.
With that $y$ noise, and keeping $N_{\rm FRB}=30,000$, we find 2-$\sigma$ confidence intervals of
$\alpha=1.37^{+0.45}_{-0.40}$ and $T_e^{(0)} = 0.95^{+0.25}_{-0.09} \times 10^6$ K, although in this case the host contribution cannot be confidently detected, as we find $b<0.23$ at 2-$\sigma$, consistent with zero.
Nonetheless, even marginal improvements over current tSZ measurements, added to information from FRBs, have the potential to constrain the thermal state of the WHIM to within tens of percent.

\subsection{Detecting HeII Reionization}

We finish with a futuristic application of our method.
The average-density IGM after reionization has a temperature $T_{\rm IGM} \approx 5,000$ K, which is kept constant by the equilibrium of photoheating due to ionizing photons and adiabatic cooling~\cite{Puchwein:2014zsa}.
It is predicted, though, that at $z\approx 3-4$ helium loses its second electron, due to the UV emission from quasars, causing additional photoheating of the IGM, and raising its temperature to $T_{\rm IGM} \sim 15,000$ K~\cite{Becker2011}.
We now explore whether this signal can be observable in tSZ, and extracted through correlations with DMs.
This is an independent test from simply searching for the expected step-like growth of DM that would occur at $z\sim3$~\cite{Zheng:2014rpa}, which requires knowledge of both DM and redshift for FRBs at those distances.

From Eq.~\eqref{eq:ywhim}, we can find the contribution to the DM of the full IGM, with $T_e \approx 15,000$ K, between $z=2$ and $z=4$, to be $y=3\times 10^{-8}$.
This is comparable with the WHIM component up to $z=0.2$, and thus  smaller than the typical tSZ that we have studied thus far.
Nonetheless, this signal is a factor of 3 larger than that detected in Refs.~\cite{Tanimura:2017ixt,deGraaff:2017byg}, where those references used $10^6$ galaxy pairs to find the minuscule contribution to the DM from filaments.
Therefore, assuming that the signal-to-noise ratio of the cross correlation scales as the square root of the number of tracers, we estimate that we would need a number $N_{\rm FRB}\approx 10^5$ of FRBs originating from $z\gtrsim 2$ to detect helium reionization, which we deem too challenging with current observatories, although it is conceivable for the distant future~\cite{Fialkov:2017qoz}.

\section{Conclusions}
\label{sec:Conclusions}

In this work we have studied the cross correlation between thermal Sunyaev-Zeldovich (tSZ) maps, and extragalactic dispersion measures (DMs).
This is an exciting new probe, which can potentially solve the mystery of the seemingly missing  baryons at low redshifts, presumed to be residing in a warm-hot intergalactic medium (WHIM). 
The tSZ effect, parametrized through the Comptonization parameter $y$, traces hot gas; whereas the DMs provide us with information about all the intergalactic gas, regardless of its temperature.
Their correlation holds, therefore, information
about regions that are colder than typically observed with tSZ maps alone, which can uncover the hidden baryons and confirm the tentative detections of a WHIM residing in galaxy filaments~\cite{Tanimura:2017ixt,deGraaff:2017byg}.

We have argued for a very direct way to perform the cross correlation between the two probes.
Given that every FRB will be roughly located within a CMB pixel,
we can assign a single value of $y$ to each FRB, given by whichever CMB pixel is co-located.
It is, then, trivial to find any correlation between the observed DM and this assigned $y$.
This correlation will be hidden below a high level of noise, which can be overcome given the expected number $N_{\rm FRB}\sim 10^4$ of FRBs to be detected per year with upcoming instruments.
We have found that this technique can be used to confidently detect the WHIM contribution to the tSZ with current CMB maps and less than 30,000 detected FRBs.
We have estimated that the cross correlation from clusters is somewhat smaller, given the relatively low crossing chance ($\sim 4\%$), and the fact that FRBs could be discarded if suspected to have transversed a cluster (either by angular localization or through the induced scattering). 
Additionally, while our simple analytic models of the thermal evolution of the WHIM are sufficient for this work, if a DM-$y$ cross correlation is detected in data it would warrant detailed studies of its origin, using simulations.

Looking into the future, if the $y$ noise can be reduced by a factor of three to ten (for instance by masking known clusters), one will not only detect the presence of the WHIM but also characterize its thermal state.
We have found that with a noise level of $y_{\rm noise}=10^{-7}$ it is possible to measure the WHIM temperature and its redshift evolution at the ten-percent level. 
Additionally, we estimate that the average host contribution can be measured at 2-$\sigma$ in this case, given our toy model.
Even a modest improvement from current data to $y_{\rm noise}=3\times 10^{-7}$ can provide a confident measurement of the shock-heated electron temperature $T_e^{(0)}$ today, as well as the approximate evolution of shock heating in the IGM. 

In summary, both tSZ maps and extragalactic DMs are powerful cosmological probes, and their cross correlation will teach us a great deal about the thermal history of the IGM. 
Given that tSZ maps already exist and FRBs will shortly be detected in the tens of thousands, a confirmation of the location of the missing baryons is imminent.

\acknowledgements

We wish to thank Liang Dai and Chema Diego for insightful comments on a previous version of this manuscript.
JBM was supported by the Department of Energy (DOE) grant DE-SC0019018.
This work was supported in part by the Black Hole Initiative, which is funded by the John Templeton Foundation.

\bibliography{FRBstSZ}{}
\bibliographystyle{bibpreferences}

\end{document}